\documentclass[twocolumn,aip,graphicx,amssymb,jmp,epsf,amsmath,superscriptaddress,reprint]{revtex4-1}

\usepackage[dvips]{color}
\usepackage{epsfig}
\usepackage{dcolumn}
\usepackage{bm}
\usepackage{setspace}

\begin{document}
\preprint{AIP/123-QED}
\title{Asymmetry in effective fields of spin-orbit torques in Pt/Co/Pt stacks}
\author{Cheow Hin Sim}
\email[Electronic address: ]{SIM$\_$Cheow$\_$Hin@dsi.a-star.edu.sg}
\affiliation{Data Storage Institute, Agency for Science Technology and Research (A*STAR), 117608 Singapore}
\author{Jian Cheng Huang}
\affiliation{Data Storage Institute, Agency for Science Technology and Research (A*STAR), 117608 Singapore}
\author{Michael Tran}
\affiliation{Data Storage Institute, Agency for Science Technology and Research (A*STAR), 117608 Singapore}
\author{Kwaku Eason}
\affiliation{Data Storage Institute, Agency for Science Technology and Research (A*STAR), 117608 Singapore}
\date{\today}%
\begin{abstract}
{Measurements of switching via spin-orbit coupling (SOC) mechanisms are discussed for a pair of inverted Pt/Co/Pt stacks with asymmetrical Pt thicknesses. Taking into account the planar Hall effect contribution, effective fields of spin-orbit torques (SOT) are evaluated using lock-in measurements of the first and second harmonics of the Hall voltage. Reversing the stack structure leads to significant asymmetries in the switching behavior, including clear evidence of a nonlinear current dependence of the transverse effective field. Our results demonstrate potentially complex interplay in devices with all-metallic interfaces utilizing SOT.}\\
\end{abstract}
\maketitle

\begin{figure}[ht]
\centering
\begin{tabular}{cc}
\epsfig{file=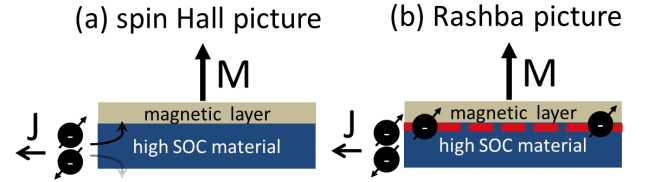,width=1.0\linewidth,clip=}
\end{tabular}
\caption{(Color online) Illustration of spin orbit torque due to (a) spin Hall effect and (b) Rashba with an in-plane current J. In the spin Hall effect, a spin current is injected in the direction shown by the up/down arrows. For the Rashba picture, an effective electric field appears at the magnetic layer/high SOC material interface.}
\label{fig:FIG0}
\end{figure}

Current-induced spin-orbit torques (SOTs) provide a novel alternative route to conventional spin-transfer torque to manipulate magnetization of a single ferromagnetic layer. Experiments have demonstrated unambiguously that in a system consisting of a bilayer interface with both a ferromagnet and a high strong spin-orbit coupling (SOC) material, stable magnetization reversal,\cite{Liu2012a, Miron2011} high frequency spin wave oscillations,\cite{Liu2012b} as well as ultrafast domain wall motion\cite{Koyama,Haazen,Ryu}can be achieved without an additional non-collinear reference ferromagnet. There are at least two mechanisms generally believed to lead to the observed SOTs, including the spin-Hall and Rashba effects. In the case of the spin-Hall effect (SHE), spin dependent scattering of a lateral electrical current generates spin currents directed towards the boundaries. The component of spin current normal to the heavy-metal (HM)/ferromagnet (FM) interface, consequently is injected into the ferromagnetic layer.\cite{Hirsch} However, in the Rashba picture, a lateral electrical current flow creates a nonlocal out-of-equilibrium spin density at the HM/FM interface due to a strong electric field gradient in film stacks having structural inversion asymmetry (SIA),\cite{Bychkov} illustrated in Fig.~\ref{fig:FIG0}. In both cases, the $\it{s-d}$ exchange interaction between these injected polarized (and/or non-equilibrium) spins and localized 3d electrons in the adjacent ferromagnet generates a SOT, equivalent to effective fields in the film plane, which acts on the magnetization.

Because SHE injects a spin-current into the ferromagnet, itinerant transversely polarized spins ($\perp$ to both $\vec{M}$ and spin polarization $\vec{p}$) scatter and accumulate at the HM/FM interface. Therefore, the dominant effective field behaves as $\vec{p}\times\vec{m}$, $\textit{i.e.}~\Delta H_{L}\propto \vec{p}\times \vec{m}$. In the Rashba picture, without spin injection, the dominant torque is expected to follow $\vec{p}$ symmetry (\textit{i.e.} $\Delta H_T\propto \vec{p}$). Thus, torque due to the transverse effective field $\it{\Delta H_{T}}$, which is orthogonal to the current flow $\vec{J}$, has symmetry similar to the classical field-like torque in magnetic tunnel junctions,\cite{Miron2010} while the spin-Hall effect, generating a longitudinal effective field $\it{\Delta H_{L}}$ parallel to current flow, leads to a torque with the same symmetry as the Slonczewski torque.\cite{Liu2012a} In a recent systematic thickness dependence study in a Ta/CoFeB/MgO stack, Kim $\it{et~al.}$ observed a transverse effective field closely associated with SHE which scales with Ta thickness.\cite{Kim} In another study, it was shown that a spin-orbital field which does not rely on the HM/FM interface can be generated non-locally in a HM/FM bilayer.\cite{Fan} These experiments indicate that more complex mechanisms may exist beyond the Rashba and SHE models. In this work, by accounting for both the anomalous Hall effect (AHE) as well as the planar Hall effect (PHE), we report measurements of the spin-orbital effective field in both the transverse and longitudinal directions in antisymmetric Pt/Co/Pt systems. We find evidence of nonlinear effects, even in such a system with all-metallic interfaces, and it is shown that only inverting the structure can alter device performance significantly. 

\begin{figure}[ht]
\centering
\begin{tabular}{cc}
\epsfig{file=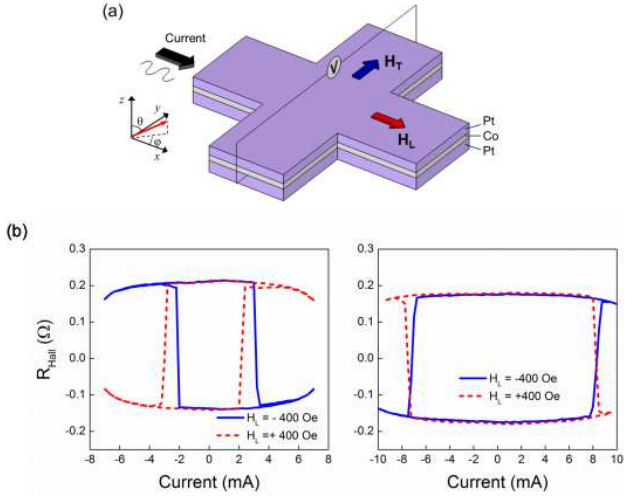,width=1.0\linewidth,clip=}
\end{tabular}
\caption{(Color online) Experimental current-induced magnetization reversal of (a) Pt 2/Co 0.6/Pt 5 (sample A) and (b) Pt 5/Co 0.6/Pt 2 (sample B) under a constant in-plane field $H_L$. The switching direction is determined by the thicker Pt layer, which corresponds to the top layer in (a) and bottom layer in (b).}
\label{fig:FIG1}
\end{figure}

\begin{figure}[ht]
\centering
\begin{tabular}{cc}
\epsfig{file=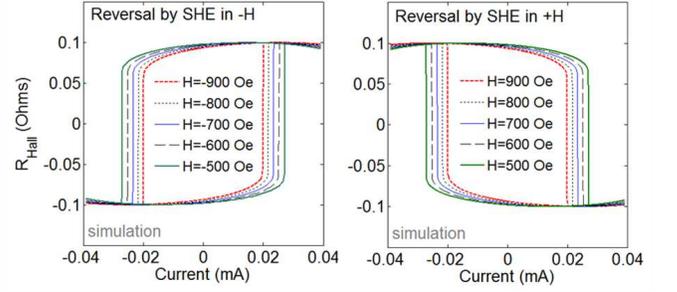,width=1.0\linewidth,clip=}
\end{tabular}
\caption{(Color online) Simulated hysteresis curves using s macrospin model for spin current injection predominantly from the bottom Pt layer in the presence of a fixed in-plane negative (left) and positive (right) field. Applying a larger field results in a smaller current needed for magnetic reversal.}
\label{fig:FIG00}
\end{figure}

We studied two film stacks consisting of, from the substrate, Pt 2/Co 0.6/Pt 5 referred to as sample A and Pt 5/Co 0.6/Pt 2 referred to as sample B (thickness indicated in nm). Such stacks with symmetrical interfaces are expected to minimize contributions from the Rashba effect, however, in some cases, relatively significant transverse fields are still observed. Owing to the difference in Pt thickness on the top and bottom layers, a net non-zero torque acts on the Co magnetization due, predominantly, to the SHE with a direction determined by the thicker Pt layer.\cite{Haazen} The films were deposited on thermally oxidized silicon wafers by dc magnetron sputtering with a base pressure of 5$\times$10$^{-9}$ Torr and patterned into 2.5$\mu$m wide Hall bars using optical lithography and Ar-ion etching. Both samples present strong perpendicular anisotropy (PMA) with an anisotropy field of 0.65T and 0.78T for sample A and B, respectively, and a longitudinal resistance R$_{xx}$ $\sim$ 880$\Omega$ in both devices. Fig.~\ref{fig:FIG1} shows the measured current-induced switching curves under a constant longitudinal magnetic field $\it{H}_{L}$ along the current direction for devices A and B. The perpendicular magnetization $\it{M_{z}}$ of the Hall bar was detected by measuring the Hall resistance, $\it{R_{Hall}}$. As shown in Fig.~\ref{fig:FIG1}, the direction of current and field determines the polarity of $\it{M_{z}}$ in switching, in agreement with the model of SHE torque in the form of $\hat{m}\times(j\times \hat{z})\times \hat{m}$, where $\hat{m}$ is unit vector in the direction of the magnetization, $\it{j}$ is the unit vector in the current density direction and $\hat{z}$ is the normal to the HM/FM interface.\cite{Liu2012c} Fig.~\ref{fig:FIG00} shows simulated hysteresis curves using a macrospin model including SHE spin-transfer torque. The  predicted symmetries are in agreement with our measured device curves shown in Fig.~\ref{fig:FIG1}.  
Under positive current and field, sample A switches from +$\it{M_{z}}$ to -$\it{M_{z}}$. Upon reversing $H_{L}$, the same switching order requires reversal of the current. This is inline with the SHE picture, as the role of the external field is to break symmetry in the energy barrier altering the gradient. Reversed fields lead to opposite energy landscape gradients, and this consequently changes the sign of the required torque for reversal. Thus, an opposite current is needed when the field is reversed. The behavior is the same, but inverted in sample B. Our experimental results also agree with previous studies of SHE induced switching,\cite{Haazen,Avci} verifying that opposite net spin injection can be achieved by tailoring the top and bottom heavy metal thickness.

To measure the magnitude and direction of the spin-orbital effective fields $\Delta H_{L}$ and $\Delta H_{T}$, we have performed lock-in measurements of the first ($\it{V_{\omega}}$) and second ($\it{V_{2\omega}}$) harmonics of the hall voltage $\it{V_{H}}$. This measurement technique was first introduced by Pi $\it{et~al.}$\cite{Pi} and subsequently used by many groups to mainly analyze HM/FM/oxide systems.\cite{Kim,Garello,Emori} When an alternating current is applied in the film plane, oscillating effective fields of the form $\it{\Delta H_{L}}sin\omega t$ and $\it{\Delta H_{T}}sin\omega t$ are generated through spin-orbit interaction which modulates the tilt of the magnetization at the drive frequency. In our measurements, $\it{V_{H}}$ generally contains anomalous Hall and planar Hall contributions, given by $\it{V_{H}=I\Delta R_{AHE}cos\theta +I\Delta R_{PHE}sin^{2}\theta sin2\phi}$, where $\it{\Delta R_{AHE}}$ and $\it{\Delta R_{PHE}}$ are the AHE and PHE saturation resistance variations; $\theta$ and $\phi$ are the polar and azimuthal angles, respectively.\cite{Lindemuth} In most previous works, PHE contribution has usually been ignored, however, its significance in the effective field evaluation has been pointed out recently.\cite{Garello,Hayashi} By performing a full cycle sweep of the in-plane field directed transverse ($\it{H_T}$) or parallel ($\it{H_L}$) to the current flow at each AC voltage $\it{V_{in}}$, the effective vector fields can be extracted as:\cite{Hayashi}
\begin{equation}
\Delta H_{L(T)}=-2\frac{B_{L(T)}\pm 2\xi B_{T(L)}}{1-4\xi^2}
\label{eq:Heff}
\end{equation}
In~(\ref{eq:Heff}), $B_{L(T)}=\frac{\partial V_{2\omega}}{\partial H_{L(T)}}\Big/\frac{\partial^2 V_{\omega}}{\partial H^2_{L(T)}}$ and $\xi=\frac{\Delta R_{PHE}}{\Delta R_{AHE}}$. The $\pm$ sign corresponds to $\hat{m}$ pointing along $\pm z$ direction.

\begin{figure}[ht]
\centering
\begin{tabular}{cc}
\epsfig{file=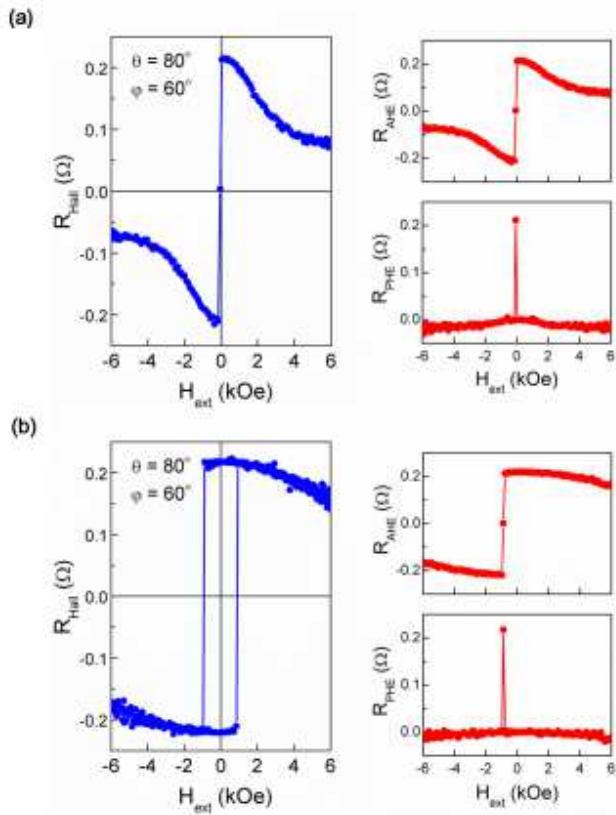,width=1.0\linewidth,clip=}
\end{tabular}
\caption{(Color online) Hall resistance $R_{Hall}$ as a function of external field Hext applied at a tilted angle for (a) Pt 2/Co/Pt 5 (sample A) and (b) Pt 5/Co/Pt 2 (sample B) with current $\it{I}$ = 1.0mA. $R_{AHE}$ relates to the perpendicular magnetization $M_z$ while $R_{PHE}$ relates to the in-plane magnetization $M_xM_y$. }
\label{fig:PHE}
\end{figure}

To determine $\Delta R_{AHE}$ and $\Delta R_{PHE}$, we first measured  $R_{Hall}$ as a function of a chosen tilted external field with $\theta=80^\circ$ and $\phi=60^\circ$ up to 6kOe. Results are shown in Fig.~\ref{fig:PHE}. AHE and PHE contributions are separated out by the anti-symmetrization and symmetrization of $R_{Hall}$ with respect to the field, according to the procedure described in Ref. 14. For our devices, we obtained $\Delta R_{AHE}=0.43\Omega$, $\Delta R_{PHE}=0.071\Omega$ for sample A and $\Delta R_{AHE}=0.44\Omega$, $\Delta R_{PHE}=0.048\Omega$ for sample B.

\begin{figure}[ht]
\centering
\begin{tabular}{cc}
\epsfig{file=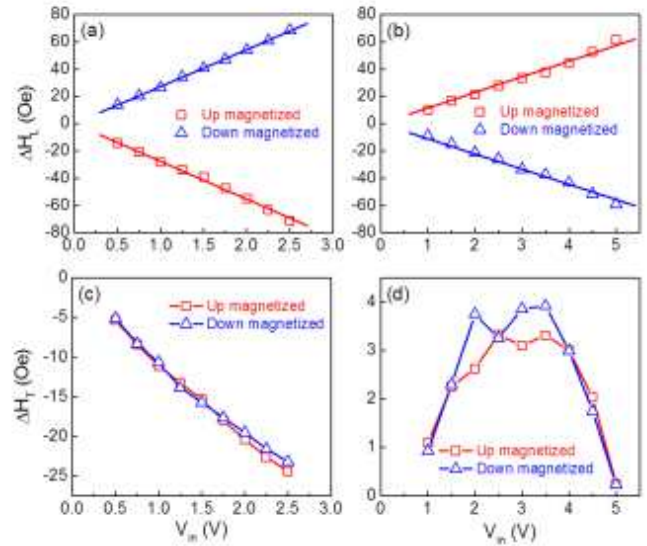,width=1.1\linewidth} 
\end{tabular}
\caption{(Color online) Input voltage ($V_{in}$) dependence of the longitudinal effective field $\Delta H_L$ for (a) sample A and (b) sample B. $V_{in}$ dependence of the transverse effective field $\Delta H_T$ for (c) sample A and (d) sample B. The triangle (blue) and square (red) symbols represent the signal when magnetization is in the down and up magnetized state, respectively.}
\label{fig:FIG3}
\end{figure}

We measured the dependence of $\Delta H_L$ on $V_{in}$, plotted in Fig.~\ref{fig:FIG3}. Due to a lower PMA in sample A (see Fig.~\ref{fig:FIG1}), the maximum sweep field ($H_{max}$=200 Oe) and input AC voltage ($V_{max}$=2.0 V) is comparatively lower than that for sample B ($H_{max}$=500 Oe, $V_{max}$=5.0 V). As expected, $\Delta H_L$ varies linearly with $V_{in}$ in both of our samples, i.e. linearly dependent on current density $J$ and its polarity depends on the $M_{z}$ direction. This is a prominent signature of the Hall effect, as the effective longitudinal field $\Delta H_L \propto J$. In addition, by inverting the structure, both magnetic as well as transport characteristics can change significantly. If we consider a uniform cross-sectional current distribution with $V_{in}$=1V corresponds to a current density $J=6.0\times 10^6$A/cm$^2$, we obtain $\Delta H_L$ of 46 Oe per $10^7$A/cm$^2$ for sample A and 19 Oe per $10^7$A/cm$^2$ for sample B from the linear fits. These values are of similar magnitude to those reported in Pt/Co/AlO$_x$\cite{Liu2012c} and Ta/CoFeB/MgO.\cite{Kim} It is interesting to note that $\Delta H_L$ is stronger in sample A. This may be attributed to its thinner bottom Pt layer which results in a rougher surface morphology and slightly degraded Pt/Co interface quality as compared to sample B.\cite{Lavrijsen} Consequently, sample A will have a weaker than expected spin current generation from the bottom Pt layer and less spin current cancellation with the top Pt layer. Assuming that $\Delta H_L$ originates solely from SHE, the net spin current will be approximately 40$\%$ of the bulk based on the drift-diffusion theory\cite{Son} and using a spin diffusion length of Pt = 1.4nm.\cite{LiuPt} We can evaluate the spin Hall angle $\theta_{SH}$ of Pt based on the relation $\Delta H_L=\frac{\hbar\theta_{SH}J}{2eM_st}$. Using $\it{t}$=0.6nm and $M_s$=1000emu/cc, we estimate $\theta_{SH}$=0.06-0.14, in agreement with previous reports for Pt.\cite{Liu_2011c, Ando, Mosendz}

$\Delta H_T$ was quantified in the same manner by sweeping $\Delta H_T(V_{in})$. Sample A shows a significantly larger transverse field vector, compared to sample B. In particular, we observe a prominent non-linear behavior of $\Delta H_T(V_{in})$ for sample B as depicted in Fig. \ref{fig:FIG3}(d), which closely resembles a quadratic dependence. The cause of the nonlinear dependence turns out to be nontrivial. \textit{A few mechanisms can be ruled out}: The non-linear behavior cannot be due to thermal effects, as they would affect both $\Delta H_T$ and $\Delta H_L$, and this is not the case. It is also not likely to be due to the Oersted field, as it is linear in $V_{in}$. In a recent article, Garello $\it{et~al.}$ highlighted that the generated transverse field in AlO$_x$/Co/Pt depends on both the current amplitude and the applied magnetic field.\cite{Garello} In their experiments, the measured torque closely follows a $sin^2\theta$ function, which is very different from the Rashba field or field-like torque of the SHE, but no explanation was given to account for it. Here, we suggest more than one possible mechanism leading to the observed nonlinear behavior: one possibility is that the $\it{s-d}$ exchange is being \textit{tuned} by the applied voltage $V_{in}$, and although the effect may normally be small, it may be comparable to a few Oe. In the past, studies of the $\it{s-d}$ exchange dependence on the itinerant electron kinetic energy (KE) has been discussed, where it was found to be proportional to the KE,\cite{myref:Stern1} which would consequently lead to a quadratic $\it J$ dependence of the field-like torque term. A second possibility is an opposition by the Rashba field. Normally, this field is considered weak, however, note that the nonlinearity appears in conditions of a weak $\Delta H_T$. The Rashba effect was, in fact, shown to be inconsequential in symmetrical stacks,\cite{Haazen,Miron2010}. However, the Rashba field could be opposing the influence from the SHE polarization in the device, observable only in the conditions of weak effective fields. Such an effect should generally appear as a weakening of $H_T$, due to opposition, which is also consistent with our measurements. The possible mechanisms are nontrivial to unambigously resolve. Nonetheless, our results clearly demonstrate that more complex mechanisms and/or interplay associated with SOTs exist, which have not been accounted for by current models. Further in-depth study is needed to identify the origin, clearly, of these non-linear SOTs to better understand how to control them in devices. 

In summary, we have investigated devices, measuring the effective spin-orbital field generated in inverted Pt/Co/Pt systems with asymmetrical Pt thicknesses. Switching was achieved by varying Pt thickness to achieve asymmetric transport, rather than the materials. The longitudinal effective field, in agreement with previous reports, linearly follows the input voltage (i.e. current density) while the transverse effective field uniquely exhibits a non-linear dependence, with very different trends between both devices studied. The observed nonlinearity cannot be explained solely by spin Hall effect, indicating the presence of other forms of spin-orbital fields. Moreover, the harmonic hall voltage measurement has been shown to be a powerful technique with good sensitivity, that enables one to unravel and understand various forms of SOTs in specially designed stacks.

The authors also wish to acknowledge very useful discussions with Dr. Tan Seng Ghee and Kong Jianfeng.
\\
%

\end{document}